\documentclass[usenatbib,usegraphicx]{mn2e}

\usepackage{subfigure}
\usepackage{longtable}
\usepackage{graphicx}
\usepackage{graphics}
\usepackage{keyval}
\usepackage{trig}
\usepackage[american]{babel}
\usepackage{url}
\usepackage{color}
\usepackage{amsmath}
\usepackage{bm}
\usepackage{longtable}

\def\beq{\begin{equation}}
\def\eeq{\end{equation}}
\def\bey{\begin{eqnarray}}
\def\eey{\end{eqnarray}}

\def\msun{M_\odot}

\def\lsim{\mathrel{\raise.3ex\hbox{$<$\kern-.75em\lower1ex\hbox{$\sim$}}}}
\def\gsim{\mathrel{\raise.3ex\hbox{$  $\kern-.75em\lower1ex\hbox{$\sim$}}}}

\def\Rdust{R_{\mathrm{dust}}}

\def\micron{\mu\mathrm{m}}
\def\dbl{D_\mathrm{bl}}

\def\Mtot{M_{\mathrm{tot}}}
\def\Mmid{M_{\mathrm{mid}}}
\def\fmax{f_{\mathrm{max}}}
\def\rmin{r_{\mathrm{min}}}
\def\rmax{r_{\mathrm{max}}}
\def\fpr{f_{\mathrm{PR}}}
\def\fdet{f_{\mathrm{det}}}

\newcommand\Eq[1]{Eq.~(\ref{#1})}
\newcommand\Fig[1]{Fig.~\ref{#1}}
\newcommand\Tab[1]{Table~\ref{#1}}
\newcommand\Sec[1]{Sec.~\ref{#1}}

\title{Steady-state evolution of debris disks around solar-type stars}
\author[]
{N. Kains$^{1,2}$ \thanks{email:nkains@eso.org}, M.C. Wyatt$^{3}$, J.S. Greaves$^{1}$\\ \\
$^{1}$SUPA, School of Physics and Astronomy, University of St. Andrews, North Haugh, St Andrews, KY16 9SS, United Kingdom\\
$^{2}$European Southern Observatory, Karl-Schwarzschild Stra\ss e 2, 85748 Garching bei M\"{u}nchen, Germany\\
$^{3}$Institute of Astronomy, University of Cambridge, Madingley Road, Cambridge CB3 0HA, United Kingdom\\
}

\begin{document}

\date{Accepted ... Received ... ; in original form ...}

\pagerange{\pageref{firstpage}--\pageref{lastpage}} \pubyear{2011}

\maketitle

\label{firstpage}
\begin{abstract}
We present an analysis of debris disk data around Solar-type stars (spectral types F0-K5) using the steady-state analytical model of \cite{wyatt07model}. Models are fitted to 
published data from the FEPS \citep{meyer06} project and various GTO programs obtained with MIPS on the \textit{Spitzer Space Telescope} at 24$\micron$ and 70$\micron$, and compared to a previously published analysis of debris disks around A stars using the same evolutionary model. We find that the model reproduces most features found in the data sets, noting that the model disk parameters for solar-type stars are different to those of A stars. Although this could mean that disks around Solar-type stars have different properties from their counterparts around earlier-type stars, it is also possible that the properties of disks around stars of different spectral types appear more different than they are because the blackbody disk radius underestimates the true disk radius by a factor $X_r$ which varies with spectral type. We use results from realistic grain modelling to quantify this effect for solar-type stars and for A stars. Our results imply that planetesimals around solar-type stars are on average larger than around A stars by a factor of a few but that the mass of the disks are lower for disks around FGK stars, as expected. We also suggest that discrepancies between the evolutionary timescales of 24$\micron$ statistics predicted by our model and that observed in previous surveys could be explained by the presence of two-component disks in the samples of those surveys, or by transient events being responsible for the $24\micron$ emission of cold disks beyond a few Myr. Further study of the prevalence of two component disks, and of constraints on  $X_r$, and increasing the size of the sample of detected disks, are important for making progress on interpreting the evolution of disks around solar-type stars.
\end{abstract}


\section{Introduction}\label{sec:intro}

Recent surveys looking for excess emission around solar-type (F, G and K spectral types, hereafter FGK) stars have found that over 15\% of these objects are surrounded by debris disks \citep{bryden06, beichman06, moromartinmalhotra07, trilling08, hillenbrand08, greaves09}. Although these disks are believed to be analogues of the Solar System's asteroid and Kuiper belts, the infrared luminosity of most detected disks around solar-type stars is usually around 100 times higher than that of the Solar System's debris structures \citep{greaveswyatt10}. This suggests that the Solar System may have unusually sparse debris, or that the detected disks around other stars are anomalously luminous, perhaps as a result of recent collisions in their planetesimal belts, producing transient excess dust. This may be caused by a process similar to the Late Heavy Bombardment (LHB) period which occured $\sim$700 Myr after the formation of the Sun \citep{booth09}. 

Among the sample of FGK stars with detected disks, several are known to harbour extrasolar planets. Interestingly, searches for asteroid and Kuiper belts analogues around some of the known extrasolar planetary systems orbiting solar-type stars, such as $\tau$ Boo, $\pi$ Men (HD 39091), $\upsilon$ And, 55 Cnc and 51 Peg did not find detectable levels of dust \citep{bryden09, beichman06asteroid}. Larger-scale searches for a correlation between the presence of a planet and that of a debris disc \citep{kospal09, moromartincarpenter07, bryden09} also concluded that the rates of excess emission were not significantly different when comparing stars with and without planets. These results could indicate that dust luminosity has fallen to near the luminosity of the Kuiper and asteroid belts of the Solar System and that these systems are past their own LHB equivalent.  Such a system would not only not feature the transient excess dust associated with such an event, but the fact that clearing occurs as part of the process that leads to the LHB (e.g. \citealt{meyer06pp}) means that dust excess emission would have to be faint. \cite{booth09}, however, suggest that such events are rare, and an alternative explanation could be that these systems might also have been born with lower-mass planetesimal belts. Studying extrasolar systems offers an important perspective for understanding the history of the Solar System. In particular, it is essential to gain an understanding of how planetesimal belts evolve around their host stars, and which mechanisms affect their structure and their luminosity in order to be able to compare observed extrasolar systems to the Kuiper belt.

\cite{kenyonbromley10} published calculations of the formation of icy planets and debris disks at 30-150 AU around 1-3$\msun$ stars and found observations of A- and G-type stars to favour models with small ($\sim$ 10km) planetesimals. \cite{carpenter09} analysed the evolution of circumstellar dust around solar-type stars using the model of \cite{dominikdecin03}, extended by \cite{wyatt07model}. This simple analytical model of steady-state debris disk evolution was developed by considering the collisional grinding down of planetesimal belts. Here we also use the model of \cite{wyatt07model} to carry out an analysis of the statistics of excess emission around FGK stars. The basic features of the model are recalled in \Sec{sec:theory}. In \Sec{sec:modelling}, we outline our modelling procedure. In \Sec{sec:data}, we briefly describe the $24$ and 70$\micron$ data we use, taken from the studies of \cite{trilling08}, \cite{hillenbrand08} and \cite{beichman06}. In \Sec{sec:fit}, we fit this data with our model and discuss our results in \Sec{sec:discussion}, comparing them to the results of the analogous study that was carried out for A stars by \cite{wyatt07Astars}. We also discuss the possible reasons for differences between the evolution of debris disks around FGK stars and that of their earlier-type counterparts, and consider individual systems that are not well fitted by our models. We conclude by looking at the implications of our models for the properties of disks around FGK stars compared to those around A stars.

\section{Debris disk model}\label{sec:theory}

In this section we recall the main features of the analytical model developed by \cite{wyatt07model} and reprised by \cite{wyatt07Astars}, as well as the assumptions made in applying this model to populations of optically thin debris disks. We start with a planetesimal belt characterised by a size distribution

\begin{equation}
\label{eq:sizedis}
n(D)\propto D^{-3.5}\, ,
\end{equation}

\noindent
where D is the diameter of the planetesimals; this is the distribution expected for a planetesimal belt in collisional equilibrium \citep{dohnanyi69}. This distribution is assumed to be valid from the largest planetesimals, of diameter $D_c$ down to the \textit{blowout} diameter $\dbl$, below which particles are blown away by radiation pressure.

For a single-radius planetesimal belt at radius $r$, with a width $dr$, the fractional luminosity of the dust emission, $f=L_{\mathrm{IR}}/L_*$ can be expressed in terms of the total cross-sectional area $\sigma_{\mathrm{tot}}$ as

\begin{equation}
\label{eq:f }
f=\frac{\sigma_{\mathrm{tot}}}{4\pi r^2}\,.
\end{equation}

Given the size distribution in \Eq{eq:sizedis}, $\sigma_{\mathrm{tot}}$ is proportional to $\Mtot$ through a constant that depends on $\dbl$ and $D_c$ \citep{wyatt07model}.

Assuming the dust particles act as blackbody emitters, the blowout diameter is given by

\begin{equation}
\label{eq:dbl}
\dbl=0.8\,\frac{L_*}{M_*}\, \frac{2700}{\rho}\, ,
\end{equation}

\noindent
where $L_*$ and $M_*$ are in solar units, and $\rho$ is the density of the dust particles in kg m$^{-3}$.
The temperature $T$ of the dust at radius $r$ is given by

\begin{equation}
\label{eq:temp}
T=278.3\,L_*^{0.25}r^{-0.5}\,.
\end{equation}

For blackbody emitters, we also know that emission from the disk at wavelength $\lambda$ is

\begin{equation}
\label{eq:fnudisk}
F_{\nu, disk} = 2.35\times10^{-11}B_{\nu}(\lambda, T)\,\sigma_{\mathrm{tot}}\,d^{-2}\, ,
\end{equation}

\noindent
where $d$ is the distance to the star in $pc$ and $F_{\nu}$ is in Jy if the Planck function $B_{\nu}$ is in Jy sr$^{-1}$.

Considering a collisional cascade in which the population within a given size range is being destroyed in collisions with other members of the cascade, but is being replenished by fragmentation of larger objects, the collision of the larger objects in the cascade solely determine the long-term evolution of the population. Hence the long-term evolution depends of the collisional lifetime $t_c$ of the planetesimals in the disk with size $D_c$. The expression for $t_c$ can be expressed as 

\begin{equation}
\label{eq:tc}
t_c = \frac{3.8\rho r^{3.5}(dr/r)D_c}{M_*^{0.5}\Mtot}\bigg\{  \frac{8}{9\,G(X_c)}  \bigg\}\, ,
\end{equation}

\noindent
where $t_c$ is in Myr, $dr/r$ is the fractional width of the disk, $D_c$ is in km, $\Mtot$ is in units of $M_\oplus$, $e$ and $I$ are the eccentricities and inclinations of the planetesimals' orbits, and the factor $G(X_c)$ is defined in \cite{wyatt07model}; note that this simplified expression is only valid when $e=I$ (which we will assume throughout this paper). $X_c$ is defined as $X_c=D_{cc}/D_c$, where $D_{cc}$ is the diameter of the smallest planetesimal that has sufficient energy to destroy a planetesimal of size $D_c$. This factor can be calculated
from the value of the dispersal threshold $Q_D^*$, which is defined as the specific incident energy required to catastrophically destroy a particle. It follows \citep{wyattdent02} that with $Q_D^*$ given in 
$\mathrm{J\,kg^{-1}}$, 

\begin{equation}
\label{eq:xc}
X_c=1.3\times10^{-3} \left( \frac{Q_D^*rM_*^{-1}}{2.25\,e^2} \right)^{1/3}.
\end{equation}

Ignoring non-collisional processes, the time-dependence of the disk mass can be worked out by solving the differential equation $d\Mtot/dt = -\Mtot/t_c$, yielding

\begin{equation}
\label{eq:massevolution}
\Mtot(t) = \Mtot(0)/[1+t/t_c(0)].
\end{equation}

As noted by \cite{wyatt07model}, the mass of the disk at $t\gg t_c$ does not depend on the initial disk mass, since $t_c(0)$ depends on $\Mtot(0)$. As a result, at a given age, there is a maximum mass that can remain after collisional evolution, and therefore a maximum infrared luminosity $\fmax$, given by \citep{wyatt07model}

\begin{equation}\label{eq:fmaxfull}
\fmax = \bigg[ \frac{10^{-6}r^{1.5}(dr/r)}{4\pi M_{*}^{0.5}t_{\mathrm{age}}}\bigg] \bigg\{\frac{4}{3\,G(X_c)}\bigg\} \left(\frac{\dbl}{D_c}\right)^{-0.5} \, ,
\end{equation}

\noindent
for $\dbl$ in $\micron$.

For observations limited by a calibration limit expressed as a flux ratio $R_{\mathrm{det}}(\lambda)=F_{\nu, \mathrm{disk}}/F_{\nu, \mathrm{phot}}$, the corresponding detection limit in terms of fractional dust luminosity, $\fdet$, is given by

\begin{equation}\label{eq:fdet}
\fdet=6 \times 10^{9}\, R_{\mathrm{det}} r^{-2} L_* T_*^{-4} B_{\nu}(\lambda, T_*)/ B_{\nu}(\lambda, 278.3 L_*^{0.25} r^{-0.5})\, .
\end{equation}

This is the threshold value we use to determine which model radii are detectable at given wavelengths in the rest of this paper.
Finally, the limit at which Poynting-Robertson (PR) drag becomes important, i.e. when radiation and gravitational timescales become similar for the smallest particles, is given \citep{wyatt07model} by

\begin{equation}\label{eq:fpr}
\fpr=50 \times 10^{-6}\, (dr/r)\sqrt{M_*/r}\, .
\end{equation}

That is, if a star system has a flux $f<\fpr$, then it is likely that it will be affected by PR drag, which could mean that in such a disk the dust component of the disk becomes spatially separated from the planetesimal belt.

\section{Modelling procedure}\label{sec:modelling}

As done by \cite{wyatt07Astars} for A stars, we apply this model to debris disk populations assuming that all stars have a planetesimal belt which is undergoing collisional processes as described by the above equations. This means that the initial state of the belts is completely determined by the parameters $r, dr, \rho, \dbl$, $\Mtot(0), L_*$ and $M_*$ (of which the latter two determine $\dbl$ through Eq. \ref{eq:dbl}), while their evolution also depends on $Q_D^*, e, D_c$ and $I$. We also set 
$\rho=2700\mathrm{kg\, m^{-3}}, dr=r/2$ and $e=I=0.05$. Values of $Q_D^*$ and $D_c$ were then constrained by fits to observational data of debris disks. 

Values of $\Mtot(0), r, L_*$ and $M_*$ were drawn from distributions of properties of stars and debris disks. We used a power law distribution for disk radii, $N(r)\propto r^{\gamma}$, and treated $\gamma$ and the limits of the radius distribution, $\rmin$ and $\rmax$, as free parameters. 

The distribution of disks at both wavelengths is affected by strong observational bias, especially for solar-type stars. For these stars, disks with radii larger than $\sim 50$ AU can only be detected at $24\micron$ if they have a large fractional excess (see e.g. Fig. \ref{fig:hist_rad} and \ref{fig:f_rad}), meaning that there is a bias towards the detection  of disks with radii lower than $\sim50$ AU, and this must be accounted for by a satisfactory model. Therefore we constrained $\gamma$ by comparing the fit of the observed distribution to a fit to the subsample of the model population which could be detected at 24 and $70\micron$ according to \Eq{eq:fdet}, rather than a fit to the whole model population. This comparison also yielded constraints on $\rmin$ and $\rmax$, although these parameters are also sensitive to the fraction of disks detected in different age bins. 

For the distribution of initial dust disk masses $\Mtot(0)$, we use the results of \cite{andrewswilliams05, andrewswilliams07}, who derived a lognormal distribution of dust masses centred on a value $\Mmid$, with a standard deviation of 0.8 dex, from submillimetre observations of protoplanetary disks in Taurus-Auriga, i.e. for $\sim$ solar-mass stars.  We use their value for the distribution's dispersion but fit $\Mmid$ as a free parameter. We choose this approach because the submillimetre data of these studies do not detect all sizes of dust grains.

\begin{figure}
  \centering
  \includegraphics[width=8cm, angle=0]{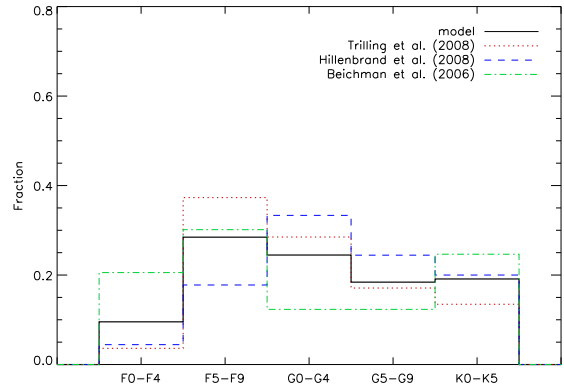}
  \caption{The distribution of spectral types for the model population, obtained by combining the observed distributions from the different surveys. The individual distributions are also plotted in red dotted line (Trilling et al. 2008), blue dashed line (Hillenbrand et al. 2008) and green dash-dot line (Beichman et al. 2006). \label{fig:dist_spectral}}
\end{figure}

A spectral type was drawn on the range F0 - K5, chosen to correspond to the range observed by the various \textit{Spitzer} programmes \citep{trilling08, hillenbrand08, beichman06}. The distributions of spectral types observed in these programmes were used as distributions for generating our model population; the resulting model distribution is plotted on \Fig{fig:dist_spectral}. We then determined a value of $\dbl$ for each spectral type, using corresponding values of $L_*$ and $M_*$ and \Eq{eq:dbl}. We also assigned values to other parameters using the relevant distributions, determining the initial conditions of each disk. It should be noted that this approach assumes that the mass and radius of the disk are independent both of each other and of the properties of the star around which they are located. 

\begin{figure}
  \centering
  \includegraphics[width=8cm, angle=0]{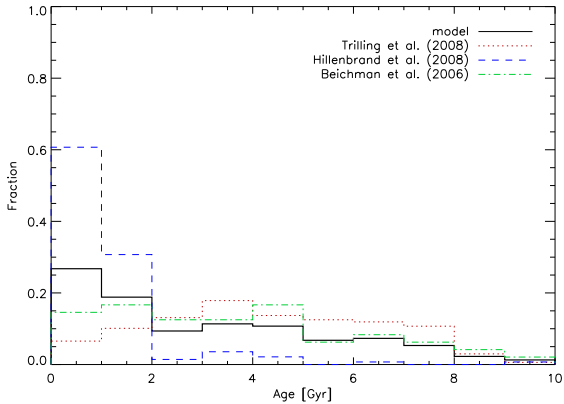}
  \caption{The distribution of stellar ages for the model population, obtained by combining the observed distributions from the different surveys. \label{fig:dist_age}}
\end{figure}

\begin{figure}
  \centering
  \includegraphics[width=8cm, angle=0]{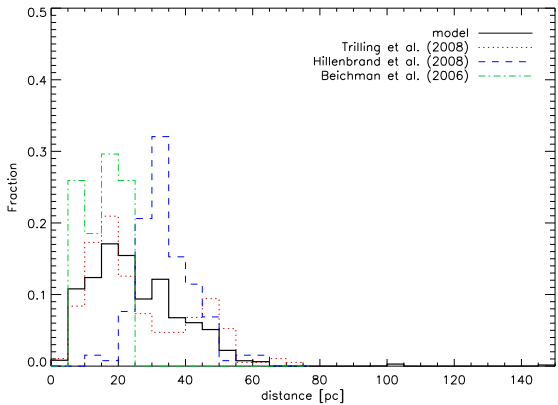}
  \caption{The distribution of distances to the stars for the model population, obtained by combining the observed distributions from the different surveys. \label{fig:dist_dist}}
\end{figure}

The systems were then evolved using Equations \ref{eq:tc} and \ref{eq:massevolution}, and values were drawn at random for the age and distance of each system, in order to determine its current ``observed" properties. Distributions for these properties were also chosen to reflect the the data sample we used (Fig. \ref{fig:dist_age}-\ref{fig:dist_dist}). The range of values for age and distance were between 0 and 10 Gyr, and between 0 and 150 pc respectively, with most stars between 0 and 60 pc (chosen to cover the same ranges as the observational data used in this paper). We do not take into account any correlation between age and distance, but although we assign a distance to each system, its value has no impact on the flux statistics on which the analysis presented in this paper is based, because all observations are assumed to be calibration-limited; we do not treat the distribution of upper limits (see Sec. \ref{sec:data}). This left $Q_D^*, D_c, \gamma, \Mmid, \rmin, \rmax$ as free parameters of the model. Finally, a good-fit model was only retained if the parameter values were realistic, i.e. consistent with ranges of values predicted by models of catastrophic collisions \citep{benzasphaug99} and planet formation \citep{kenyonbromley02}.

\section{Data}\label{sec:data}

A list of all stars with $70\micron$ excess emission only in the sample we use in this paper is given in \Tab{tab:dat_70}, while a list of sources that show excess at both $24$ and $70\micron$ is given in \Tab{tab:dat_2470}; stellar properties for these are published values resulting from Kurucz model fits, or in cases for which these were not available, were computed using Schmidt-Kaler relations \citep{schmidt-kaler82}. Using the $24-70\micron$ colour temperature of the dust, and assuming that the dust acts as a blackbody, we also calculated a disk radius for the disks in \Tab{tab:dat_2470} with \Eq{eq:temp}. We do not include the debris disks published recently by \cite{koerner10} because they do not report stellar ages or non-detections in their data. This sample would only add one disk detected at both $24$ and $70\micron$, and therefore would not influence our results significantly. 

We made a cut in stellar age, excluding objects with an age below $30$ Myr in order to avoid protoplanetary disks affecting our statistics. We also excluded disks for which 1-$\sigma$ error bars were above a threshold $\sigma_{t}=5$ mJy. This threshold was chosen empirically to avoid having noisy flux upper limits being wrongly counted as large excesses. These cuts affected mostly the FEPS data \citep{hillenbrand08}, with 6 disks detected at both wavelengths being removed from their published sample. The reason for this is the difference in the observational approaches of the SIMTPF and FEPS observations: the observations of \cite{trilling08} and \cite{beichman06} at $70\micron$ are calibration-limited while those of \cite{hillenbrand08} are sensitivity-limited; at $24\micron$, all observation are calibration-limited (see \Tab{tab:dat_2470}). The consequence of this is that most of the FEPS disks are very bright disks, since these are the only ones for which good photometry could be obtained.

The resulting sample consists of 46 disks detected at $70\micron$ only and 17 detected at both $24$ and $70\micron$. Below is a short description of each subsample we used.

\subsection{Trilling et al. (2008) sample}

\cite{trilling08} collected a sample of 193 F, G and K stars observed with the Multiband Imaging Photometer for \textit{Spitzer} (MIPS). Together with data already published by \cite{bryden06}, the whole sample gives a view of debris disks around stars with masses and ages similar to that of the Sun, covering a range of spectral types between F0 and K5. Their ``solar type" sample was selected using criteria on the spectral type (F5-K5) and luminosity class (IV or V), as well as setting a minimum photospheric $70 \micron$ flux and signal-to-noise (S/N) ratio. From the combined FGK sample, 27 show significant excess at $70 \micron$, meaning an excess with $\chi_{70} \geq 3$, where $\chi_{70}$ is defined as

\begin{equation}
\label{eq:chi70}
\chi_{70} = \frac{F70 - P70}{\sigma_{70}}\, ,
\end{equation}

where $F70$ is the total $70 \micron$ flux measured, $P70$ is the predicted photospheric flux at $70 \micron$, and $\sigma_{70}$ is the error bar associated with the flux measurement. On top of these 27 objects, 3 more are classified as excess sources for reasons detailed in \cite{trilling08}, but one is removed due to the stellar age not being determined, bringing the total number of excess objects in the sample to 29. Finally, a re-reduction of data for HD 101259 \citep{su10} found that this system did not in fact have significant excess, so we removed this object from the sample. This left 28 objects with excess emssion, including 6 systems for which significant excess flux is detected at both $24$ and $70 \micron$.


\begin{table*}
  \caption{List of sources with 70$\mu m$ excess only. Flux ratios and limits are given as total flux divided by photospheric flux. $R24$ is defined as $R24=F24/F24_*$, and similarly, $R70=F70/F70_*$. $C24$ and $C70$ are the $3-\sigma$ calibration limits at $24$ and $70\micron$, defined as $C_{\lambda}=1+3\sigma_{*, \lambda}/F_{*, \lambda}$.}
  \begin{tabular}{cccccccc}
\hline
Star name		& Sp. type		& $d$	&$t_{age}$	&$R24$	 &C24	&$R70$	 &C70\\
			& 			& $(pc)$	&$(Gyr)$		&	 &$(3\sigma)$& &$(3\sigma)$\\

\hline
Trilling et al.\\
\hline
HD 1581&	F9 V &		8.6&	3.02&	   0.98&   	1.10&        1.4&    1.3   \\
HD 3296&	F5	&		47&	2.5&		   1.02&	1.10&	4.8&    2.8 \\
HD 17925	&	K1 V&		10&	0.19&	    1.05&	1.10&	3.4&   1.7   \\
HD 19994	&	F8 V&		22&	3.55&	      1.00&	1.10&	1.7&  1.5\\
HD 20807	&	G1 V&		12&	7.88&	     1.04&	1.10&	1.8&   1.5\\
HD 22484	&	F8 V&		17&	8.32&	      1.02&	1.10&	1.9&  1.3\\
HD 30495	&	G1 V&		13&	1.32&	  1.06&	1.10&	5.5&  1.6    \\
HD 33262	&	F7 V&		12&	3.52&	     1.03&	1.10&	1.7&   1.4  \\
HD 33636	&	G0	&		29&	3.24&	     1.00&	1.10&	7.0&    2.2 \\
HD 50554	&	F8 V&		31&	4.68&	     1.00&	1.10&	8.4&   3.4  \\
HD 52265	&	G0	&		28&	6.03&	     0.99&	1.10&	4.8&  2.9 \\
HD 57703	&	F2	&		44&	2.3&		     1.03&	1.10&	9.3&   3.3  \\
HD 72905	&	G1.5 V&		14&	0.42&	     1.06&	1.10&	2.3&  1.5  \\
HD 75616	&	F5	&		36&	4.8&		     1.03&	1.10&	10&   2.5  \\
HD 76151	&	G3 V&		11&	1.84&	     1.03&	1.10&	2.4&    1.6   \\
HD 82943	&	G0	&		27&	4.07&	     1.02&	1.10&	17&   3.1  \\
HD 110897&	G0 V&		17&	9.7&		    0.98&	1.10&	4.3&   1.9 \\
HD 115617&	G5 V&		8.5&	6.31&	     1.04&	1.10&	4.0&   1.5 \\
HD 117176&	G5 V&		18&	5.37&	    0.98&	1.10&	1.8&   1.3 \\
HD 128311&	K0	&		167& 0.39&	    0.94&	1.10&	3.0&   2.3      \\
HD 206860&	G0 V&		18&	5.00&	    1.03&	1.10&	2.3&  1.5 \\
HD 212695&	F5	&		51&	2.3&		     1.00&	1.10&	9.5&  3.3  \\

\hline
Hillenbrand et al.\\
\hline
HD 6963&		G7 V &			27&		1.00&	  1.05&     1.13&     13&       8.5  \\
HD 8907&		F8 &				34&		0.32&	  1.05&     1.13&    46 &       12   \\
HD 31392&		K0 V &			26&		1.00&	  1.02&     1.12&     20&       8.5   \\
HD 35850&		F7/8 V &			27&		0.03&	    1.14&     1.14&     4.9&       3.9\\
HD 38529&		G8 III/IV &			42&		3.16&	    0.96&    1.12&     4.3&        3.1  \\
HD 72905&		G1.5 &			14&		0.10&	   0.84&    1.10&     2.7&          2.1\\
HD 122652&		F8  &				37&		3.16&	  1.08&     1.13&     24&        10  \\
HD 145229&		G0  &			33&		1.00&	  1.09&     1.14&     20&        9.1  \\
HD 150706&		G3 V &			27&		1.00&	   1.05&     1.13&     8.7&      6.4	\\
HD 187897&		G5 &				33&		1.00&	  1.03&     1.12&     14&       7.5   \\
HD 201219&		G5 &				36&		1.00&	  1.07&     1.13&     18&       11   \\
HD 209253&		F6/7 V &			30&		0.10&	  1.14&     1.14&     14&       6.8   \\

\hline
Beichman et al.\\
\hline
HD 38858&		G4 V &		16&	4.57&	1.00&		1.30&	10&		3.0\\
HD 48682&		G0 V &		17&	3.31&	1.00&		1.59&	12&		2.0\\
HD 90089&		F2 V &		21&	1.78&	1.00&		1.59&	2.3&		1.7\\
HD 105211&		F2 V &		20&	2.53&	1.01&		1.19&	11&		2.4\\
HD 139664&		F5 IV-V &		18&	0.15&	1.10&		1.24&	5.9&		1.6\\
HD 158633&		K0 V &		13&	4.27&	0.90&		1.53&	18&		2.0\\
HD 219623&		F7 V &		20&	5.06&	1.04&		1.12&	3.0&		1.7\\

\hline
\hline

  \end{tabular}
  \label{tab:dat_70}
\end{table*}



\begin{table*}
  \caption{List of sources with 24 and 70$\micron$ excesses. Flux ratios and limits are given as total flux divided by photospheric flux. $R24$, $R70$, $C24$ and $C70$ as defined as in \Tab{tab:dat_70}. Stellar properties are either published values found by fitting Kurucz models to the data when these were available, or calculated using Schmidt-Kaler relations for main sequence stars.}
  \begin{tabular}{cccccccccccccc}

\hline
Star name		& Sp. type		& $d$	&$t_{age}$	& $L_*$	&$M_*$	&$f$		&$r$		&$f/f_{max}$	&$f/\fpr$	&$R24$	 &C24&$R70$	 &C70 \\
		& 		& $(pc)$	&$(Gyr)$ &$/L_{\bigodot}$ &$/M_{\bigodot}$	&$(/10^{-5})$		&(AU)		&	&	&	 &$(3\sigma)$	 &	 &$(3\sigma)$\\

\hline
Trilling et al.\\
\hline
HD 166&		K0 V &		14&	5.0& 		0.42&	0.79&		5.9&		        9.1&            1.99&        8.0&         1.14&	1.10&      	6.9&    	1.8 \\
HD 3126&	F2	&		42&	3.5&		2.9&		1.5&			13&			21.8&	1.86&	20&       1.16&	1.10&	27&    	3.3  \\
HD 10647&	F9 V&		17&	6.3&		1.8&		1.1&			34&			21&		5.97&	59&           1.21&	1.10&	51&  		 2.1	\\
HD 69830&	K0 V&		13&	4.7&		0.42&	0.79&		20&			1.0&		 836&	9.0&          1.47&	1.10&	1.5& 		1.5\\
HD 105912&	F5	&		50&	1.8&		3.2&		1.4&			7.9&			7.7&		5.71&	7.4&          1.52&	1.10&	11&  		3.0\\
HD 207129&	G0 V&		16&	5.8&		1.5&		1.1&			12&			15.3&	 3.42&	18&           1.17&	1.10&	16&  		2.8\\

\hline
Hillenbrand et al.\\
\hline
HD 25457&		F7 V &		19&		0.10&	2.1&		1.3&		10&		17&		0.05&      14.8&    1.31&     1.16&     18&       5.0   \\
HD 37484&		F3 V &		60&		0.10&	3.5&		1.5&		32&		17&		0.23&      43.5&    1.43&     1.29&     46&       14   \\
HD 85301&		G5  &		32&		1.0&		0.71&	0.92&	13&		8&		1.64&      14.8&    1.36&     1.17&     13&        8.4  \\
HD 202917&		G5 V &		46&		0.03&	0.66&	0.92&	25&		9&		0.08&      31.0&    1.63&     1.20&     28&       16   \\
HD 219498&		G5 &			150&	0.32&	4.9&		0.92&	20&		31&		0.11&      46.4&    1.29&     1.15&     25&       14   \\

\hline
Beichman et al.\\
\hline
HD 25998&		F7 V &		21&	0.6&		2.4&		1.2&		4.5&		13&		0.27&      5.8&		1.14&	1.12&	4.2&		2.2\\
HD 40136&		F1 V &		15&	1.3&		4.3&		1.6&		1.9&		5.9&		2.24&       1.5&		1.13&	1.12&	1.6&		1.5\\
HD 109085&		F2 V &		18&	1.3&		2.9&		1.5&		15&		5.9&		14&       11.8&		1.99&	1.12&	5.9&		1.6\\
HD 199260&		F7 V &		21&	3.2&		2.4&		1.2&		3.3&		14&		0.96&       4.4&		1.11&	1.12&	3.5&		2.0\\
HD 219482&		F7 V &		21&	6.1&		2.4&		1.2&		3.6&		18&		1.04&       5.5&		1.08&	1.12&	4.4&		1.7\\

\hline
\hline

  \end{tabular}
  \label{tab:dat_2470}
\end{table*}

\subsection{Beichman et al. (2006) sample}

The sample of \cite{beichman06} includes objects observed in the frame of other projects, including radial velocity search teams, coronagraphy and interferometry missions. As a result, it includes some low-mass close stars. They selected stars within 25 pc (with a few exceptions for the earlier-type stars) and excluded targets with binary companions within 100 AU on the grounds that binarity might prevent the formation or long-term stability of planetary systems. Their selected sample was then observed with MIPS at $24$ and $70\micron$ and four stars were observed further with the \textit{Infrared Array Camera} (IRAC) in order to help determine their photospheric flux. 

Based on the criterion expressed by \Eq{eq:chi70}, \cite{beichman06} identify 12 stars out of the 88 in their sample to have a $70 \micron$ excess at the $3\,\sigma$ level or better. Amongst these, 5 also display significant excess at $24\micron$.

\subsection{Hillenbrand et al. (2008) sample}

\cite{hillenbrand08} presented data obtained using MIPS, IRS and IRAC, with most of their systems observed at $3.6, 4.5, 8.0, 13, 24, 33, 70$ and $160\micron$, as part of the FEPS (\textit{Formation and Evolution of Planetary Systems}) program. They report 25 systems with excess flux at $70\micron$ and $\mathrm{S/N}_{70\micron} \geq 3$. Amongst this sample, 13 also have excess emission at $24\micron$. Since photospheric sensitivity at $70\micron$ could not be achieved without very long integration times except for the closest stars in the sample, the sensitivity of their observations was determined by a target detection threshold of dust emission, which they expressed relative to dust emission in a young Solar System model. For most targets, the survey was sensitive to 5-10 times the dust emission predicted by the model.

The sample of \cite{hillenbrand08} is different from those of \cite{trilling08} and \cite{beichman06}, as the former sample is age-selected, leading to an even distribution in logarithmic age bins, while the latter are volume-limited, leading to a linear age distribution. This difference is taken into account in the analysis that follows when fitting power laws to time-dependent parameters, and the data cuts we made were chosen to minimise the effect of poorly constrained photospheric fluxes in the FEPS sample on our modelling of the statistics.

\section{Results}\label{sec:fit}

\subsection{Fit to the 70$\micron$ statistics}

\begin{figure}
  \centering
  \includegraphics[width=8cm, angle=0]{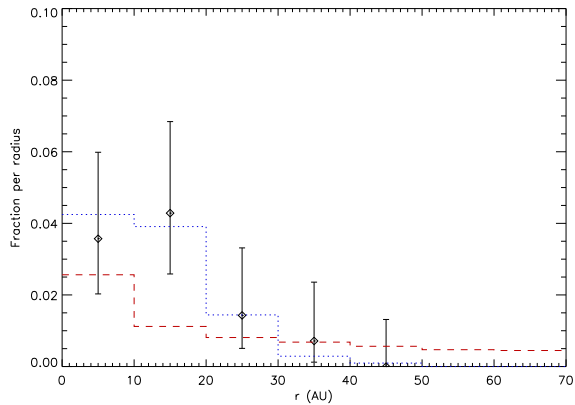}
  \caption{Histogram for the distribution of disk radii of the combined observational samples (diamonds) for disks detected at both 24 and 70 $\micron$, shown with error bars calculated for small samples by Gehrels et al. (1986). Also plotted are the distributions for the entire model population (red, dashed line) and the model population that could be detected at 24 and 70$\micron$ (blue, dotted line). \label{fig:hist_rad}}
\end{figure}

Following the procedure described in \Sec{sec:modelling}, we found best-fit values of $D_c=450\,\mathrm{km}$ and $Q_D^*=3700\,\mathrm{J\,kg^{-1}}$. These are consistent with models of catastrophic collisions \citep{benzasphaug99}. We also found best-fit parameters for the radius distribution of $\rmin=1$ AU, $\rmax=160$AU and $\gamma= -0.60 \pm 0.35$, similar to the value of $\gamma \sim -0.8 \pm 0.3$ found in the A stars study by \cite{wyatt07Astars}. The values of $\sigma$ were obtained using error bars on measured fluxes reported in the literature. A histogram of the distribution of disk radii for the model and observed data is shown on \Fig{fig:hist_rad}. In order to reflect our small sample sizes, we use error bars for small-number Poisson statistics as calculated by \cite{gehrels86}.  Our radius distribution extends to lower radius than the one found by \cite{lohne08}, who used a value of $\rmin=20$ AU. However they limited their data sample to G stars, and their analysis used radius values calculated by assuming realistic grain emission rather than blackbody emission. Furthermore, we argue in \Sec{sec:effective} that the true radius distribution should be larger than the one we derive. Finally, we find a best-fit value for the median of the initial dust mass distribution of $\Mmid=4 M_{\bigoplus}$, which is consistent with results for the disk mass-stellar mass relation of \cite{natta04}, who find an approximate range of total disk masses around solar-mass stars of $-2.5 < \mathrm{log}(M_{\mathrm{disk}}/M_*)<-0.5$. This value is lower than that of $\Mmid=10 M_{\bigoplus}$ found for A stars by \cite{wyatt07Astars}.

\begin{figure*}
  \centering
  \subfigure
  {
  \includegraphics[width=8cm, angle=0]{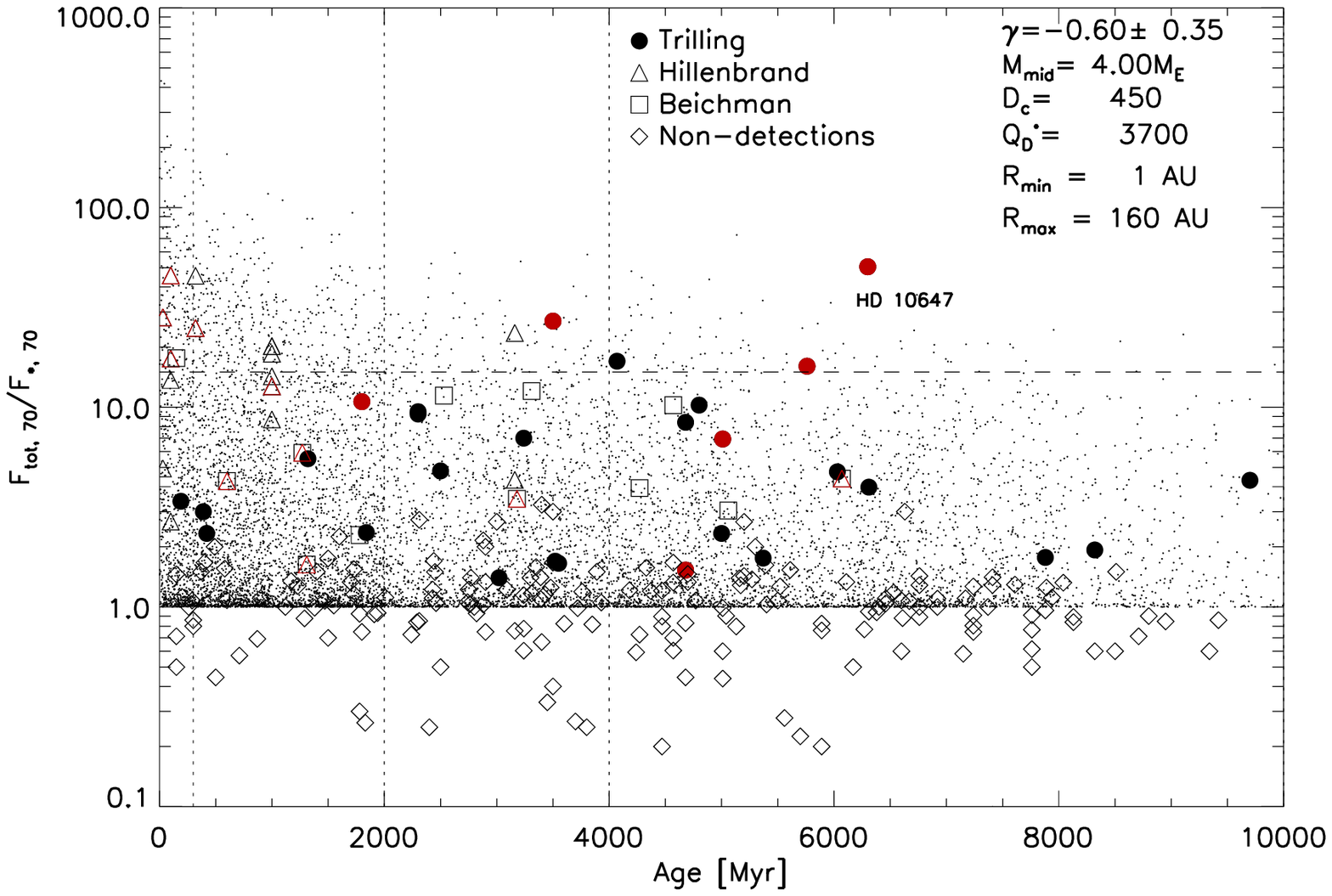}
  }
    \subfigure
  {
  \includegraphics[width=8cm, angle=0]{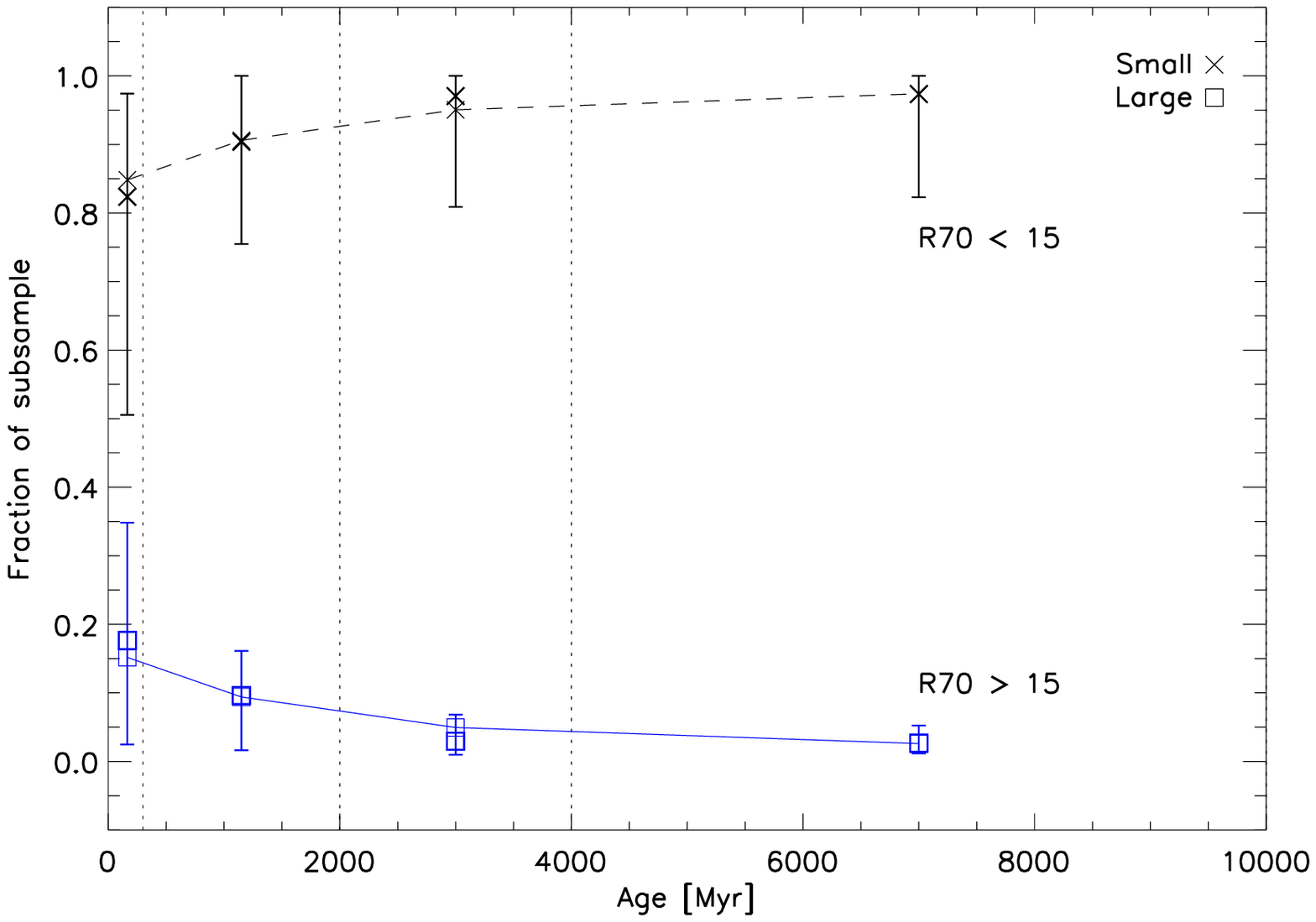}
  }
  \caption{\textit{Left:} total 70$\micron$ flux divided by the photospheric flux against age. The model population is shown with small dots, and the observations of Trilling, Hillenbrand and Beichman are shown as filled dots, triangles and squares respectively. Disks detected at $70 \micron$ with $<3 \sigma$ confidence are shown with open diamonds. Red symbols indicate disk emission detected at $>3\sigma$ at both $24\micron$ and $70\micron$. The horizontal dashed line separates the populations with small and large excess flux. \textit{Right:} fractional populations of stars with different flux ratios for different age bins ($0.03-0.3$ Gyr, $0.3-2$ Gyr, $2-4$ Gyr, $4-10$ Gyr). Small ($R_{70}=F_{70, tot}/F_{70, phot} < 15$) and large excesses ($R_{70} > 15$) are shown as crosses and squares respectively. Model predictions are connected with lines and observed values \citep{trilling08, hillenbrand08, beichman06} are plotted with small-sample Poisson statistics error bars (Gehrels et al., 1986) where appropriate. On both panels, the dotted vertical lines indicate the limits of the age bins.\label{fig:F70}}
\end{figure*}

The models found with these parameters are plotted on \Fig{fig:F70}, which show that the statistics are reproduced by the model convincingly, as they were as for A stars \citep{wyatt07Astars}; this is shown on the right-hand panel of \Fig{fig:F70}. The flux ratio shows very slow evolution past the earliest age bins, which is what is seen in the data as well. We use two excess categories, small ($R70<15$) and large ($R70>15$). The threshold value $R70=15$ was chosen empirically to avoid upper limits contaminating the sample of stars with large excess emission. The large excess fraction makes up $\sim 20\%$ of excesses at early ages, falling to a few percents within $\sim 3$ Gyr.

\subsection{Fit to the $24\micron$ statistics}\label{sec:fit24}

\begin{figure*}
  \centering
  \subfigure
  {
  \includegraphics[width=8cm, angle=0]{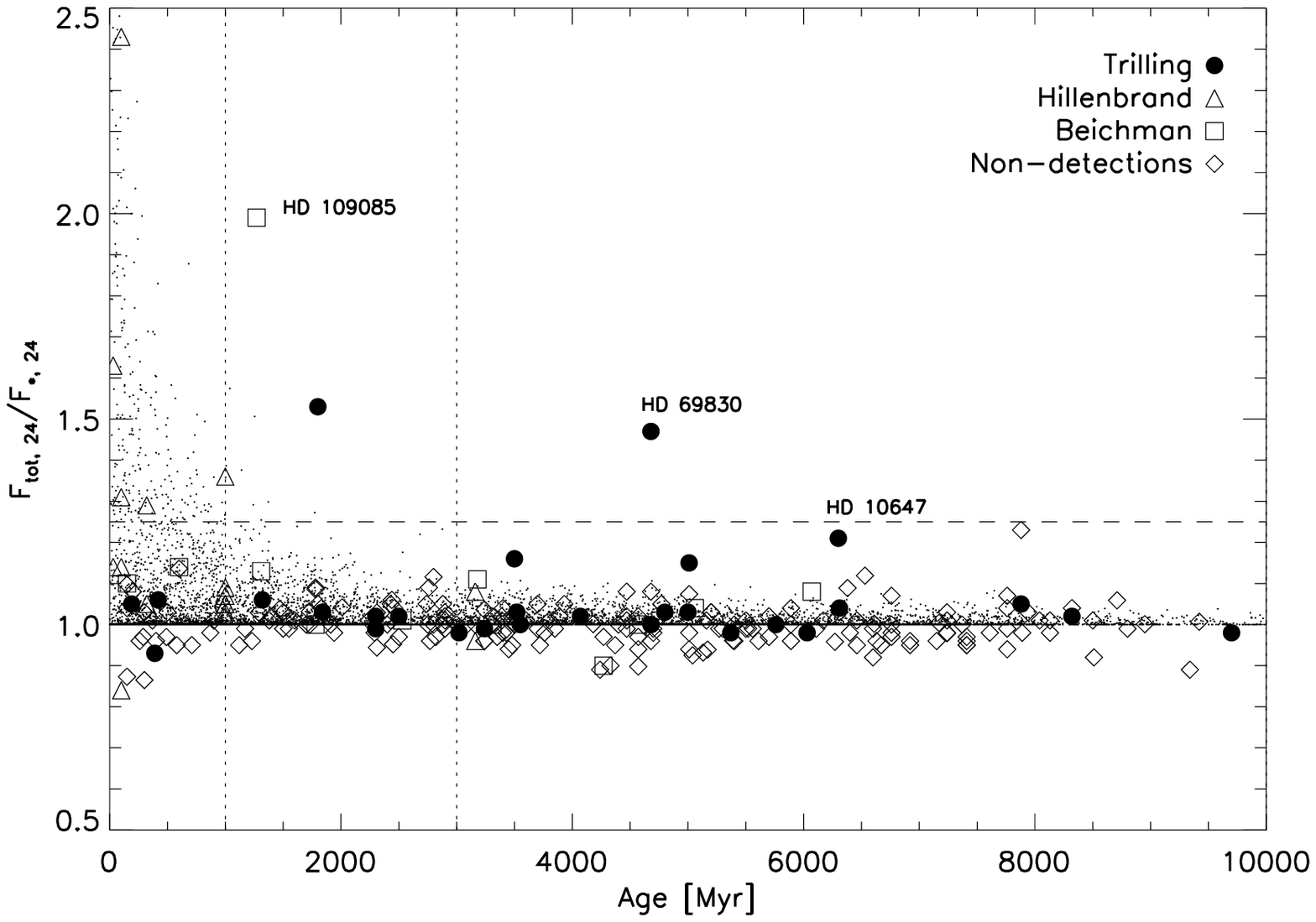}
  }
 \subfigure
 {
   \includegraphics[width=8cm, angle=0]{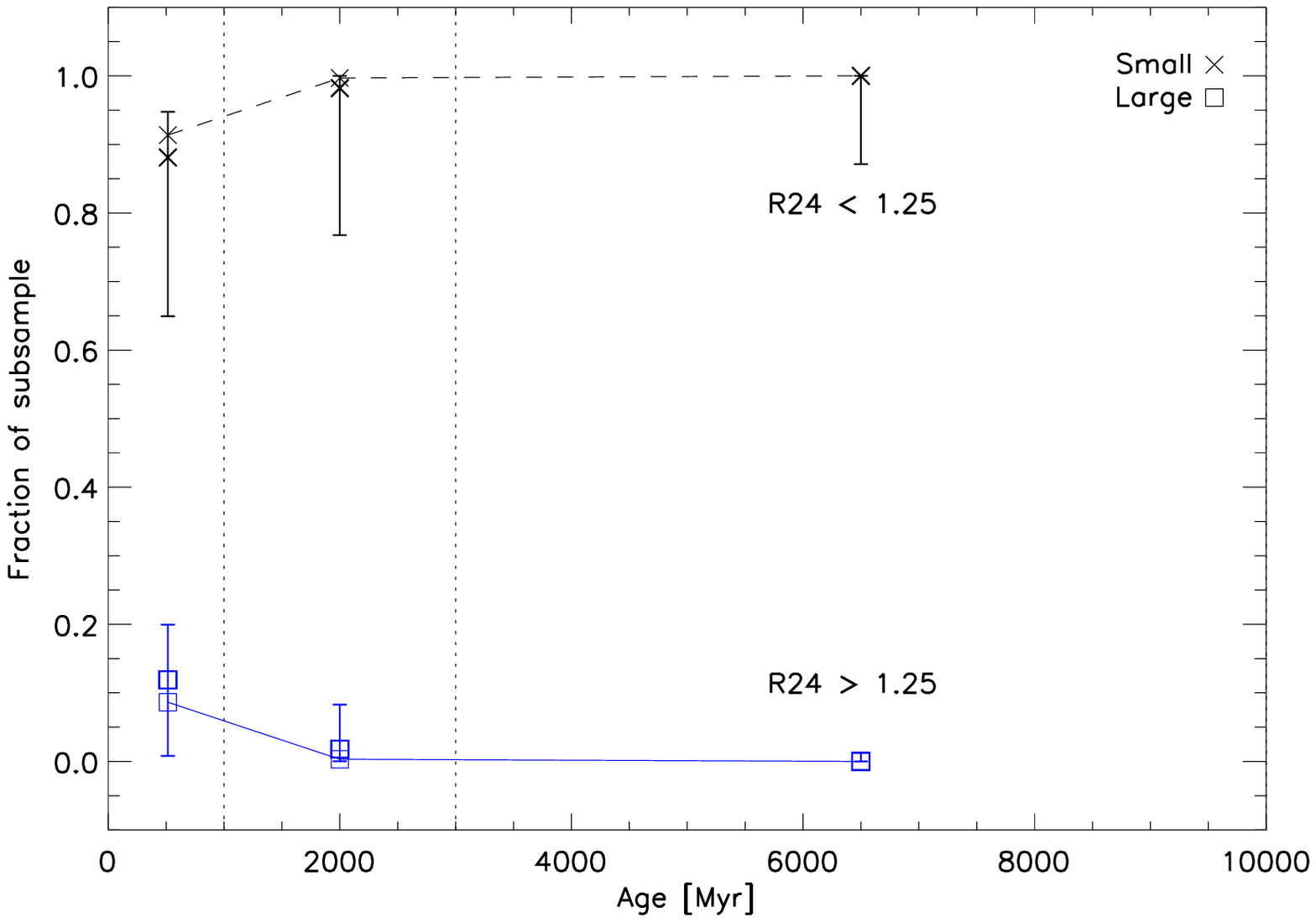}
 }
 
  \caption{\textit{Left:} total 24$\micron$ flux divided by the photospheric flux against age. Symbols are the same as for \Fig{fig:F70}. A horizontal dashed lines separates the populations with small and large excess flux. \textit{Right:} fractional populations of stars with different flux ratios for different age bins ($0.03-1$ Gyr, $1-3$ Gyr, $3-13$ Gyr). Small ($R_{24}=F_{24, tot}/F_{24, phot} < 1.25$), and large excesses ($R_{24} > 1.25$) are shown as crosses, triangles and squares respectively. Model predictions are connected with lines and observed values \citep{trilling08, hillenbrand08, beichman06} are plotted with small-sample Poisson statistics error bars (Gehrels et al., 1986) where appropriate.  On both panels, the dotted vertical lines indicate the limits of the age bins. \label{fig:F24} }
\end{figure*}

\begin{figure}
  \centering
  \includegraphics[width=8cm, angle=0]{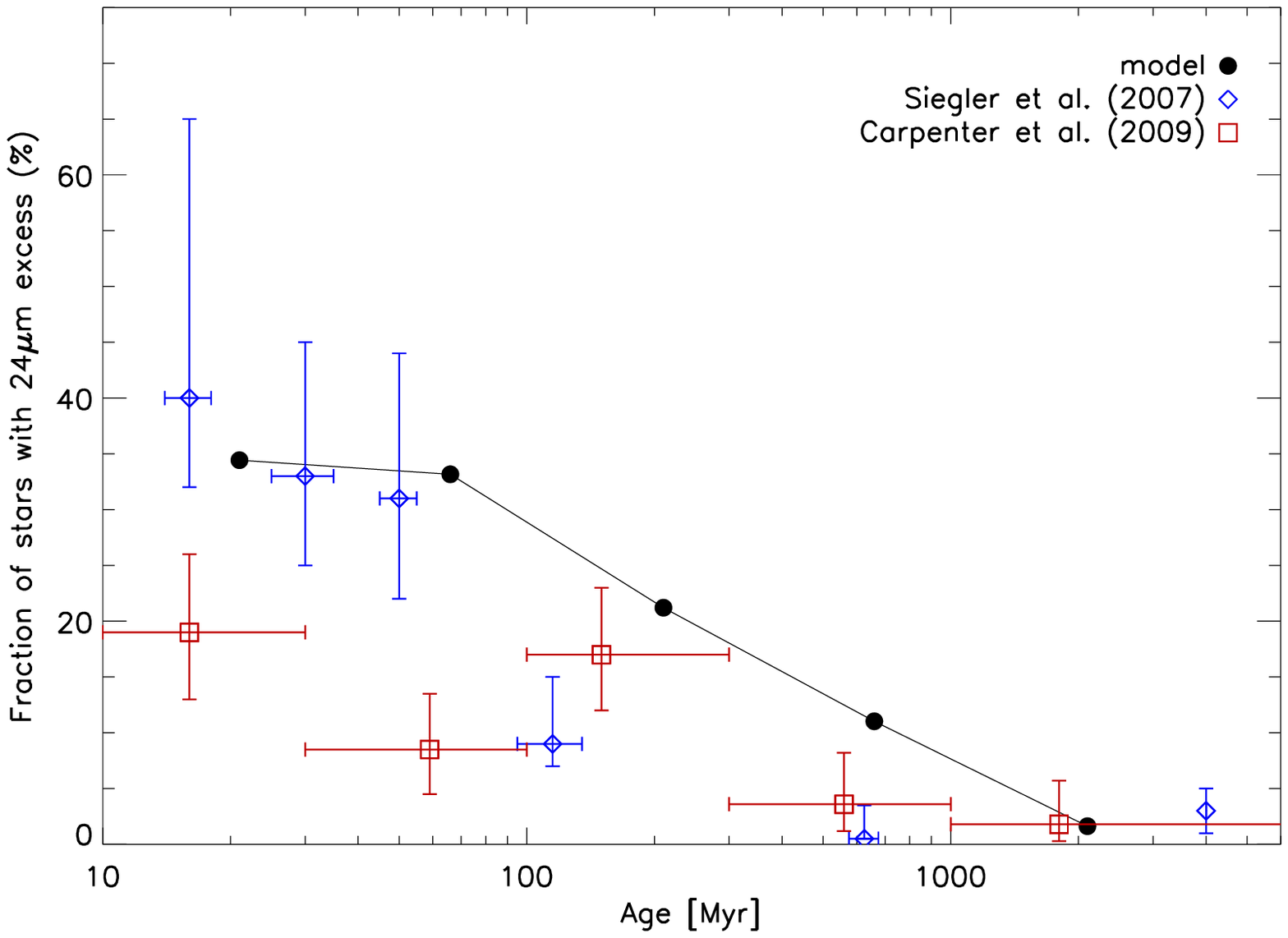}
  \caption{Time evolution of the fraction of stars with 24$\micron$ excess for the best-fit model (black solid line and filled circles), compared with the findings of Siegler et al. (2007, blue diamonds) and Carpenter et al. (2009, red squares). The plot indicates that our model evolves on a longer timescale as found by these studies, although it is in agreement with the results of Siegler et al. (2007) for young ($<100$ Myr) systems and Carpenter et al. (2009) for older ones. Note that for these studies, an excess was defined as $R24 > 1.15$, so we adopted this threshold for this plot. \label{fig:excess_age}}
\end{figure}

\Fig{fig:F24} shows the model fit (with the parameter values given in the previous section) to the observed $24\micron$ statistics. The statistics both in the model and the data samples show that $24\micron$ excess flux evolves on a timescale of $\sim 1.0-1.5$ Gyr, similar to what was found by \cite{lohne08}, but slower than what is seen in the observations of \cite{siegler07} and \cite{carpenter09} (see also \Fig{fig:excess_age}).  $24 \micron$ statistics are well fitted by the model and are particularly crucial in constraining the parameters which determine the evolution timescale of the models ($Q_D^*, e, D_c, \Mmid$), as they show clearer time evolution than the 70$\micron$ data; this is clearly visible on the left panel of \Fig{fig:F24}. 

\Fig{fig:excess_age} shows the temporal evolution of the fraction of systems in the model population that have excess emission at 24$\micron$. Also plotted for comparison are the data presented by \cite{siegler07} and \cite{carpenter09} (the latter an extension of the results of \citealt{meyer08}). Our model is within 1-2$\sigma$ of the results of \cite{siegler07} for early-age ($<100$ Myr) systems, as well as those of \cite{carpenter09} for systems with age $>100$ Myr, but the temporal evolution we predict is significantly slower than that seen by those surveys. We discuss possible reasons for this in Section \ref{sec:timeevolution}. 

The combined fit to the 24 and 70$\micron$ statistics has a goodness-of-fit of $\chi^2/\mathrm{d.o.f.}=1.05$.

\subsubsection{Discussion of individual sources}\label{sec:individual}

From \Fig{fig:F24}, it is clear that a number of systems are not well fitted by our models. We discuss shortly the systems labeled on the plot: HD 109085, HD 69830 and HD 10647. 

HD 109085 was found by \cite{sheret04} to have a radius of $\sim$180 AU, compared to the blackbody radius of 6 AU used in this analysis. This system was also reported to have 2 separate disk components: a component with $r<$3.5 AU (recent work also shows that for this component $r>0.5$ AU, e.g. \citealt{smith09}) and a cold component at $\sim$150 AU \citep{smithwyattdent08}, with the middle region possibly cleared by a planetary system \citep{wyattgreaves05}. Our results suggest that the hotter disk component is transient, whereas the cold component is evolving in steady state. HD 69830 has 3 known planetary companions \citep{lovis06}, as well as an asteroid belt around 1 AU from the star \citep{beichman05, lisse07}. Since a disk with such a small radius is expected to have processed its material at 4.7 Gyr (the age of HD 69830), this system can be considered anomalous; fitting it in our models would require unrealistic parameter values. 

HD 10647 also has a known Jupiter-mass planetary companion and a large cold disk at a radius of $\sim 300$ AU has been detected at sub-millimetre wavelengths \citep{liseau08}. \cite{lawler09} fit the IRS spectrum with a disk radius of 16-29 AU, while \cite{liseau10} imaged this disk with Herschel, finding a peak in the dust emission at 85 AU. This system has the largest 70$\micron$ excess in our sample. As for HD 69830, finding a model that fits this object would require extreme parameter values, as it is unusually bright compared to other disks of similar age and colour temperature. This could be caused by the disk around HD 10647 having unusually strong planetesimals, or having been recently stirred.

\subsubsection{Model predictions of $24\micron$ excess detection rates}

We compared the detection rates of excess emission at $24\micron$ predicted by our models to each of the three surveys used in this analysis. Since at $70\micron$ the large scatter of excesses and accuracy of flux measurements makes determining an excess threshold difficult and excess is quantified instead with the $\chi_{70}$ statistic (Eq. \ref{eq:chi70}), we only ``predict" detection rates at $24\micron$. We use the relevant age, spectral type and distance distribution of each survey to estimate their detection rates from the generated model population. For the survey by \cite{trilling08}, we find a predicted detection rate of 3.3\%, in excellent agreement with their detection rate of $4.2^{+2.0}_{-1.1}$\%. We also find a good agreement with the results of \cite{beichman06}, who found a detection rate of 7.3\%, while our model predicts 7.1\%. Finally, the study of \cite{hillenbrand08} constrained the $24\micron$ rate to be $<40$\%, with our model predicting a rate of $16.9$\%.

\section{Discussion}\label{sec:discussion}

\subsection{Significance of best-fit parameters}\label{sec:significance}

The best-fit values we find for $Q_D^*$ and $D_c$ are significantly different from those found for the models of \cite{wyatt07Astars} for debris disks around A stars. Both $Q_D^*$ and $D_c$ are an order of magnitude larger (values for these parameters in the A stars study are 150 $\mathrm{J\,kg^{-1}}$ and 60 km respectively). This could suggest either that our models are incomplete and these parameter values compensate for the value of one of several other free parameters being poorly chosen or fitted, or that the properties of debris disks around later-type stars are actually different. Other regions of parameter space were also explored, and our best fit constitutes the best trade-off between the $\chi^2$ goodness-of-fit statistic for the evolution of excess rates (\Fig{fig:F70} and \ref{fig:F24}), and realistic parameter values, as stated in \Sec{sec:modelling}. 

In comparison with the study of A stars, we find different values for $\rmin, \gamma, \Mmid, D_c$ and $Q_D^*$. The most significant differences are the higher values of $Q_D^*$ and $D_c$, although the exact values of the individual free parameters are poorly constrained and are not as informative as combinations of these parameters. In this case, the values point to a slower evolution of $F_{70, tot}/F_{70, phot}$ and of $F_{24, tot}/F_{24, phot}$ compared to what was found for disks around earlier-type stars (e.g. \citealt{su06}). It should also be emphasised that while FEPS data is available at several wavelengths for some of the objects presented here, we only consider data at 24 and 70$\micron$. Using different wavelengths colours might result in different radii, temperatures and luminosities, for example if there is cool dust present in the disk. \cite{carpenter09} indeed found that the dust temperatures derived from MIPS 24 and 70$\micron$ were lower than those inferred from spectra obtained with IRS (\textit{Infrared Spectrograph}), and \cite{hillenbrand08} find different colour temperatures when using 33 and 24$\micron$ or 70 and 33$\micron$ flux density ratios.

$Q_D^*$ might vary because the composition of the disk may be different around later-type stars, or the bodies making up the disk may be more compacted by previous collisions, making them stronger and accounting for a higher intrinsic strength of the disks. The value of $D_c$ indicates an initial population made of large objects, up to Pluto-size asteroids rather than smaller planetesimals, hinting at Kuiper-belt-like properties. However, the difference between solar-type and A stars results may also be due to inaccuracies in deriving the radius of disks with the $24-70\micron$ colour temperature. 

Because small grains are inefficient emitters, and we are assuming blackbody radiation from all grains in the disk, the actual radii may be larger than the derived values. This is confirmed by results of disk imaging which show that disks are generally 2-3 times larger than the blackbody value (e.g. \citealt{maness09}). We therefore emphasise that in this paper we are finding an \textit{effective} value for the parameters $Q^*_D$ and $D_c$, as was done for the A stars study. The discrepancy in effective values between A and FGK stars could then be caused by an dearth of small grains around A stars compared to solar-type stars. This is because the blowout size of grains is larger around earlier-type stars ($\sim$ 10 $\micron$ for A stars and $\sim 1 \micron$ for a G0 star), meaning that with larger number of inefficient emitters, the emission is \textit{less} well described by blackbody emission for disks around FGK stars. Hence the ratio of real radius to blackbody radius could be larger for solar-type stars than for A type stars. Despite \textit{effective} values being different, the real values of the planetesimal strength and largest planetesimal size may therefore be similar for solar-type and A stars. In the next section, we discuss and quantify this.

\subsection{Effective and real parameters}\label{sec:effective}

\cite{bonsorwyatt10} used grain modelling, taking into account the non-blackbody nature of the grains and a realistic size distribution, to show that for A stars, real disk radii are expected to be larger than blackbody radii by a factor of $\sim 2$ (with some dependence on radius and spectral type), in agreement with observations of resolved disks. They also derived scaling laws between realistic and effective values for parameters within the context of the modelling presented here. Assuming that the real radii are larger by a factor $X_r$ compared to blackbody radii, then the real parameters can be calculated from parameters which were found using the blackbody assumption, by making sure that all disks have the same luminosity evolution when the radii are changed. In the following discussion we use a simplified notation so that $M \equiv \Mmid$, $Q \equiv Q_D^*$ and $D \equiv D_c$. Furthermore, we use the subscript $r$ to denote a real value, while its absence denotes an effective value: $Q_r$ refers to the real planetesimal strength, while $Q$ refers to the fitted (effective) parameter.

Disks start off with the same luminosity if

\begin{equation}\label{eq:effective_dc}
M_r\, D_{r}^{0.5}\,X_r^{-2} = M\, D^{0.5} \, ,
\end{equation}

\noindent
from Eq. 3-4 of \cite{wyatt07Astars} (see also Eq. 16 of that paper), and disks evolve on the same timescale if

\begin{equation}\label{eq:effective_qd}
Q_{r}^{5/6}\, D_{r}^{0.5}\, X_r^{7/3} = Q^{5/6}\, D^{0.5}\, .
\end{equation}
 
Note that these constraints are for an average population, and that they can be violated on a case-by-case basis. The real parameters for each population cannot be derived from the parameters of the blackbody fit, because there are 4 free parameters and 2 constraints. However, it is possible to compare the real parameters of the FGK and A star populations from their blackbody fits. We rearrange \Eq{eq:effective_dc} and (\ref{eq:effective_qd}) to derive expressions for $X_r$ and $Q_r$ for the FGK stars population in terms of the equivalent A stars parameters. We use the additional subscripts $A$ for A stars and $F$ for FGK stars in the following comparison. 



If we assume that the strength of the planetesimals follows $Q_r \propto D_r^{1.5}$ (i.e. that the strength of the planetesimals is mainly gravitational), and that planetesimals have the same composition around A and FGK stars, we can also rearrange these equations to find that, with the parameter values found in this paper and in \cite{wyatt07Astars} ($Q_D^*=150$ Jkg$^{-1}$, $D_c=60$ km and $M=10\,M_{\bigoplus}$) as well as the value of $X_r$ found for A stars by \cite{bonsorwyatt10},

\begin{equation}\label{eq:mrdr}
\frac{M_{r, F}}{M_{r, A}} \simeq 3.4\, \frac{D_{r, A}}{D_{r, F}} \, ,
\end{equation}

\noindent
and using this to derive an expression for the scaling factor yields

\begin{equation}\label{eq:xrdr}
\frac{X_{r, F}}{X_{r, A}} \simeq 4.8\,\left(\frac{D_{r, A}}{D_{r, F}}\right)^{3/4} \,.
\end{equation}

We can expect $X_{r}$ to be larger for FGK stars than for A stars for the reason mentioned earlier that there are more inefficient emitters around FGK stars. Unfortunately the overlap between the sample of disks resolved with \textit{Spitzer} and the data sample used in this paper is too small for this to be useful to constrain $X_r$ for FGK stars, although imaged disks such as HD 181327 \citep{schneider06} suggest that a value of $\sim 3$ is a good estimate for $X_{r, F}$. Using $X_{r, A}=2$ \citep{bonsorwyatt10}, the value of $X_{r, F}$ is consistent with the estimate of $\sim 3$ that is expected from imaged disks (e.g. \citealt{schneider06}) if the planetesimals around FGK stars are larger than those around A stars by a factor of $\sim 5$. Putting this number into \Eq{eq:mrdr}, this means that the median disk mass around FGK stars is $\sim 0.72$ times that for disks around A stars. This is in agreement with the findings of e.g. \cite{natta04} that earlier-type stars are expected to have more massive disks. 

Finally, the values found for $\gamma, \rmin$ and $\rmax$ point to a slightly flatter disk distribution for FGK stars compared to their earlier-type counterparts. This, however, does not take into account the possible systematic difference in actual disk radius contained in the scaling factor $X_{r, F}$, which would result in more large disks around FGK stars to what was found for the A stars.

\subsection{Time evolution of the 24$\micron$ statistics}\label{sec:timeevolution}

As pointed out in Section \ref{sec:fit24}, the temporal evolution we derive in the model presented here is significantly slower than that observed by \cite{carpenter09} and \cite{siegler07}. The need for the model to yield slower time evolution is most obvious on Fig. \ref{fig:F24} and \ref{fig:rad_age}, the latter of which shows that several disks in the sample have small radii despite being relatively old and therefore expected to have processed all their mass. 

We suggest two possible reasons for this discrepancy. In the first, the faster evolution seen by \cite{carpenter09} and \cite{siegler07} could be due to the disks having two components. Surveys at 24$\micron$ that focus on young stars would trace the hotter component that evolves on short timescales (being at smaller radii), whereas surveys of older stars would trace the colder component. Indeed, the data we use in this paper comes from surveys looking at older stars, for which the hot component would have faded below detectable levels. In contrast, the sample of \cite{carpenter09} has over 80\% of stars younger than 1 Gyr, and \cite{siegler07} have over 70\% of young stars.

Although \cite{carpenter09} conclude from IRS spectra that 24$\micron$ excess emission arises from cold dust, the spectrum of a two-component disk could still look like a cold dust emission spectrum if the hot component, while bright enough to produce detectable $24\micron$ excess emission at young ages, is too faint to make the $24-70\micron$ colour look like hot dust emission. Therefore more evidence is needed to determine whether two-component disks are responsible for the discrepancy in observed evolutionary timescales.

The alternative explanation is that the fast 24$\micron$ evolution seen by \cite{carpenter09} and \cite{siegler07} is the true evolution of single temperature disks (e.g. Kuiper belts), and that this really does disappear by a few 100 Myr (see \Fig{fig:F24}-\ref{fig:excess_age}). The 24$\micron$ emission from the older systems that contribute to the survey statistics modelled in this paper then must arise from a transient component. This is already thought to be the case for some of the old systems for which the 24$\micron$ emission comes from a hot component within a few AU (e.g. \citealt{wyatt07Astars}), and is perhaps favoured by the fact that IRS spectra do not show evidence for two-component disks, but this in not the case for all of the older systems, such as HD10647 which would have to be a significant outlier in such a model.

Here we have modelled one of these possible cases, where $24\micron$ comes from the cold component of a two-component disk. It is difficult to favour an interpretation with the small number of disks detected at both $24$ and $70\micron$, but additional detections from future surveys will help distinguish between these possible explanations. Finally we also note that as in this paper we use a single-temperature disk model; fitting this data with an extended disk model and a range of temperatures could also significantly improve the fit. 

\subsection{Fractional luminosity vs. radius}\label{sec:f_rad}

\begin{figure*}
  \centering
   \subfigure
  {
  	\includegraphics[width=8cm, angle=0]{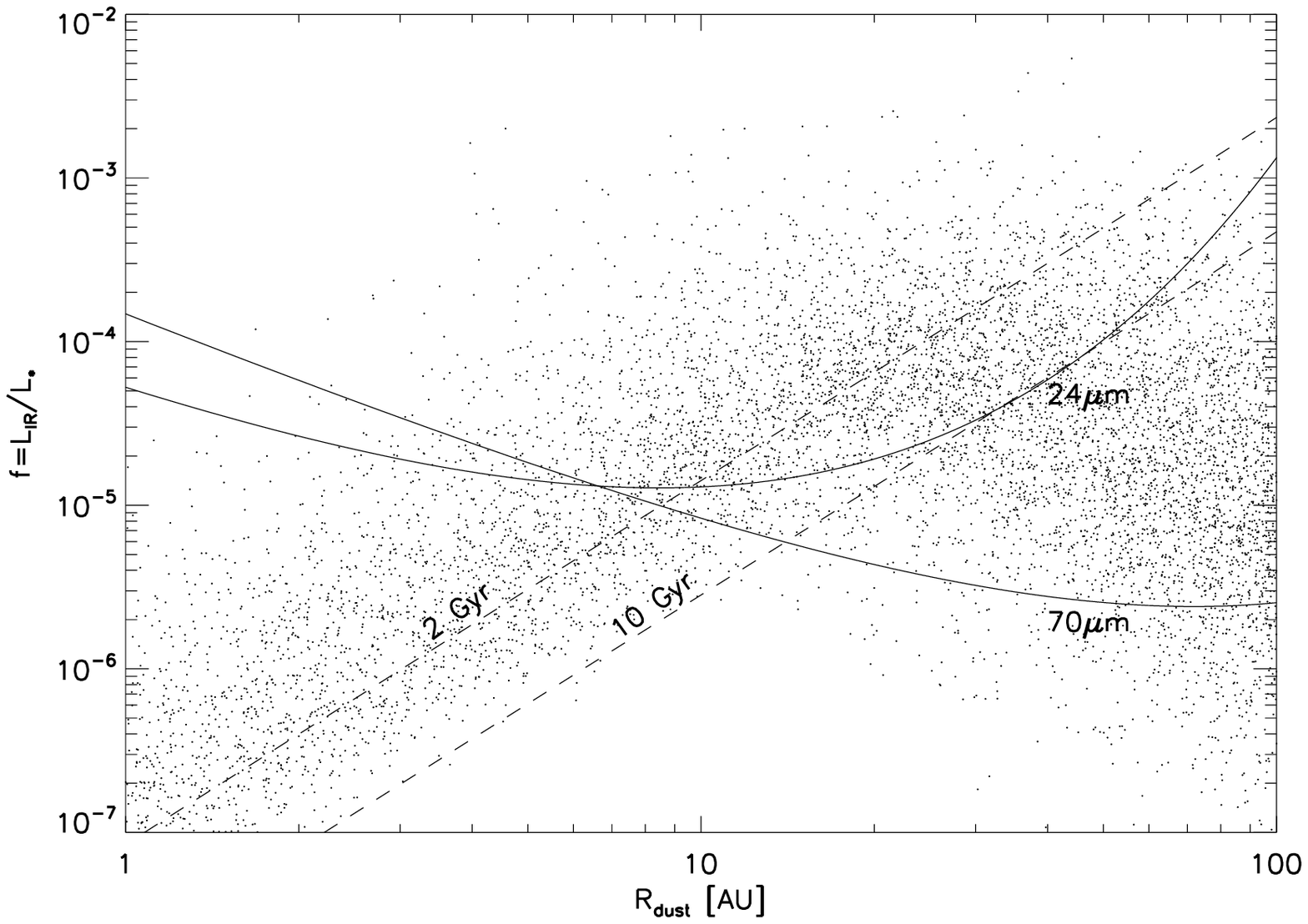}
  }
    \subfigure
  {
  	\includegraphics[width=8cm, angle=0]{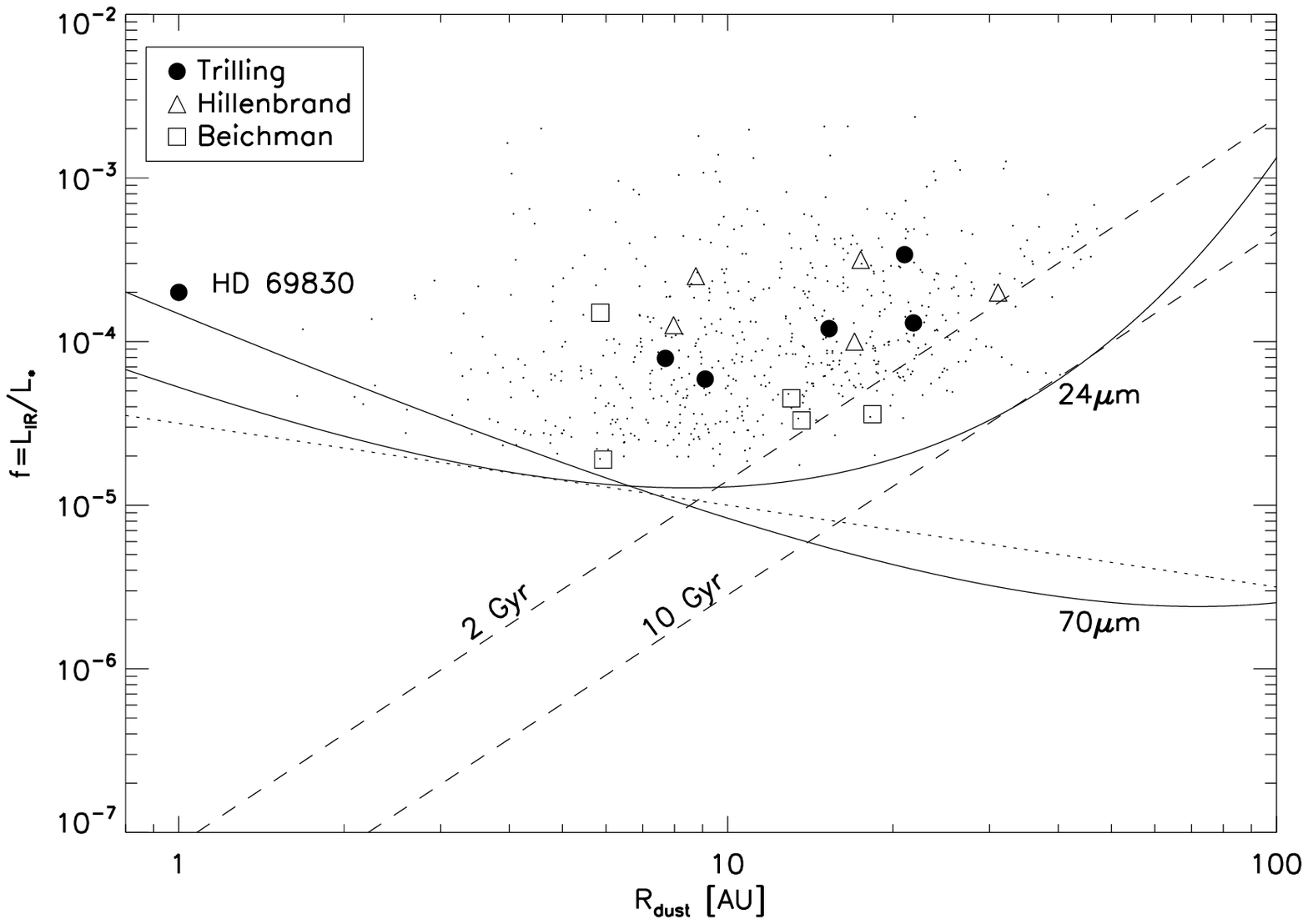}
  }
  \caption{Dust luminosity $f=L_{IR}/L_*$ plotted as a function of disk radius $\Rdust$. The left-hand panel shows the whole model population, while the right-hand panel only has the model population that could be detected at $24$ and $70\micron$. The model population is shown with small dots, and data symbols are the same as for \Fig{fig:F24}. Detection thresholds for an F0 dwarf are indicated by a solid ($70 \mu\mathrm{m}$) and a dashed ($24 \mu\mathrm{m}$) line. Lines showing the maximum possible fractional luminosity for an F0 dwarf are shown as dashed lines for 2 and 10 Gyr. The dotted line indicates the limit, given by \Eq{eq:fpr}, where Poynting-Robertson drag becomes important for an F0 dwarf. \label{fig:f_rad}}
\end{figure*}

\Fig{fig:f_rad} shows a plot of fractional luminosity $f$ against disk radius, with the observed data plotted over the model population. Also plotted are the detection limits at both $24$ and $70\micron$, given by \Eq{eq:fdet}, and the lines of maximum fractional luminosity at 2 and 10 Gyr, given by \Eq{eq:fmaxfull}, for an F0 dwarf and the best-fit model parameters. The detection thresholds are calculated assuming calibration limits of $R_{24}=0.11$ and $R_{70}=0.5$. In theory, disks should lie on or below the line of maximum luminosity for their age and spectral type, although as noted by \cite{wyatt07Astars}, the precise location of these lines depends on parameters which may vary between disks such as $Q_D^*$, $e$, $D_c$, and the spectral type of the disk's host star. The sharp increase in the $24\micron$ threshold for disks with radii larger than $\sim 50$ AU means that the disks with larger radii, i.e. those disks which are expected to have the strongest emission at $70\micron$ will be more difficult to detect at both wavelengths as they will require high fractional dust luminosity.

Values of $f/\fmax$ for disks detected at 24 and 70$\micron$ are given in Table \ref{tab:dat_2470}. Since $\fmax$ depends on the disk radius\footnote{Note that the calculation of $\fmax$ is unaffected by the discussion in \Sec{sec:effective} on the difference between real and effective radius, as long as $\fmax$ is calculated using both effective parameters and blackbody radius (which would give the same value that would be calculated from the real parameters and real radius).}, a good estimate can only be made for those disks detected at both wavelengths, so we do not include a value $f/\fmax$ for disks detected at $70\micron$ only. Several disks stand out in Table \ref{tab:dat_2470} and on \Fig{fig:f_rad}. In particular, HD 69830 \citep{trilling08} has a luminosity $f$ over 800 times the maximum theoretical value $\fmax$ for its age, radius and spectral type. This comes from a very small disk radius of 1 AU combined with a relatively old stellar age of 4.7 Gyr. A fit to the model population yields $f \propto r^{0.72}$, while a fit to the observed population of disks in the \cite{trilling08} and \cite{beichman06} samples, without anomalously bright systems, and without disks known to have two components (HD 109085), gives a relation $f \propto r^{0.68 \pm 0.05}$, in excellent agreement.

\subsection{Fractional luminosity vs. age}

\begin{figure*}
  \centering
  \subfigure
  {
  	\includegraphics[width=8cm, angle=0]{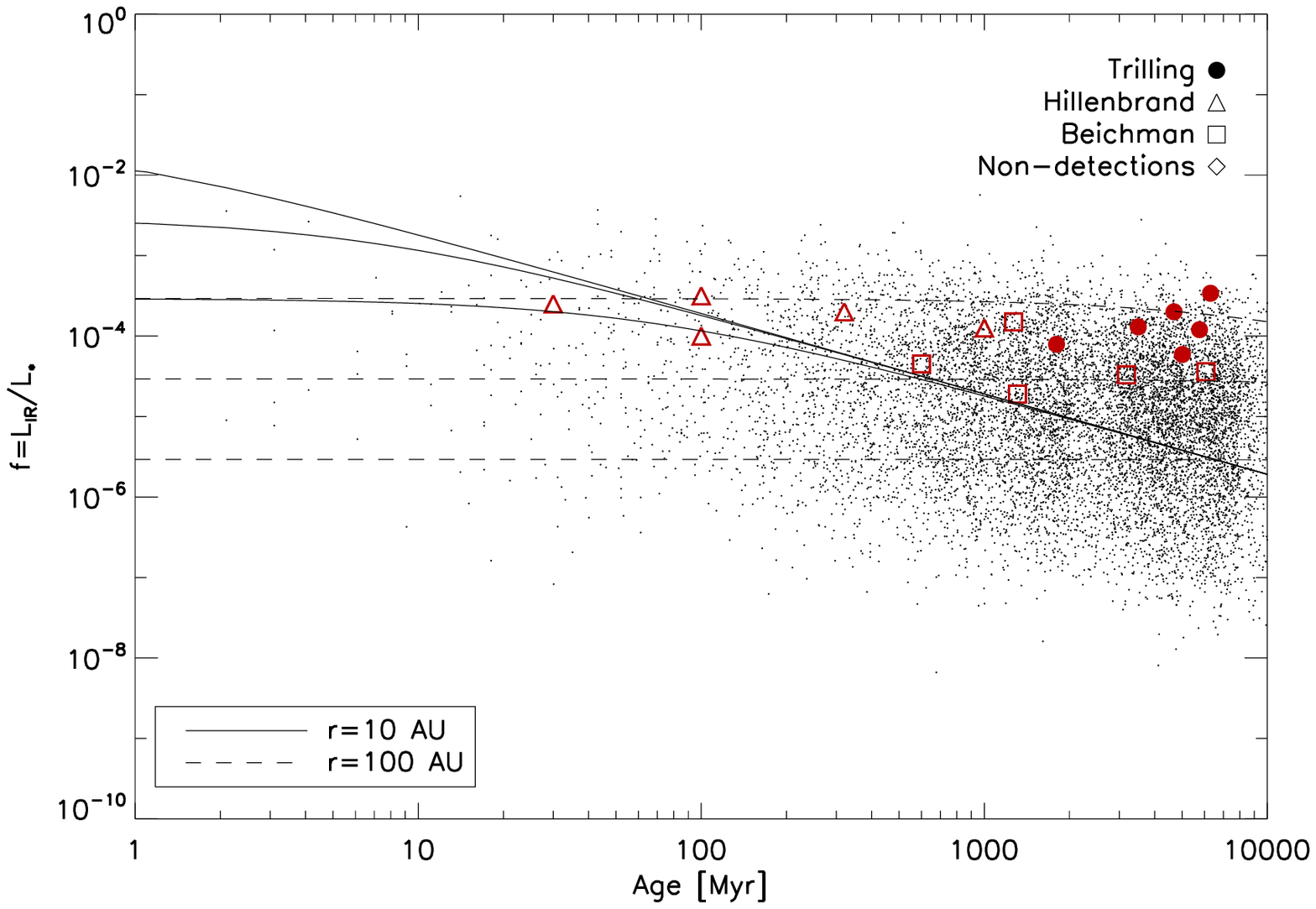}
  }
  \subfigure
  {
	\includegraphics[width=8cm, angle=0]{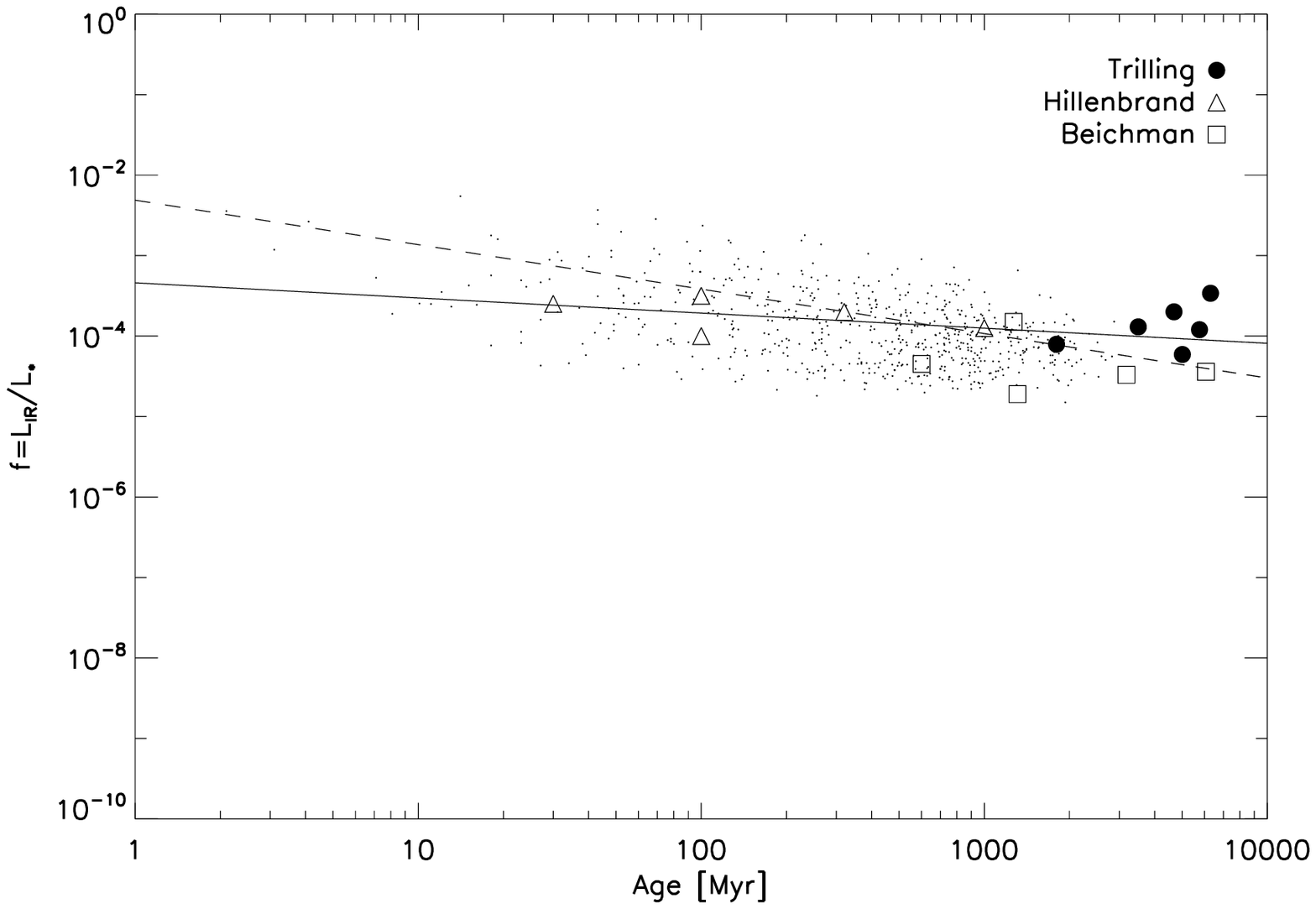}
  }
  \caption{Dust luminosity $f=L_{\mathrm{IR}}/L_*$ plotted as a function of age. The model population is shown with small dots, and data symbols are the same as for \Fig{fig:F24}. For $<3 \sigma$ detections, the fluxes plotted are maximum values. \textit{Left:} Entire model population and observed objects detected at both $24$ and $70\micron$. Also shown are theoretical evolutionary tracks for disks of radii 10 (solid line) and 100 (dashed line) AU, and for initial disk mass of 3, 30 and 300 $M_{\bigoplus}$ (with the higher masses corresponding to the lines with higher early-age fractional luminosities), around an F0 star. \textit{Right:} Same with only the model population that could be detected at $24$ and $70\micron$ plotted. Also plotted on the right-hand panel are power-law fits to the model (dashed line) and the data (solid line). The power law fit to the model population is in agreement with the fit to the data (see text). \label{fig:f_age}}
\end{figure*}

\Fig{fig:f_age} shows the fractional luminosity of disks as a function of age, as well as theoretical evolution lines for disks of initial mass 3, 30 and 300 $M_{\bigoplus}$ of radii 10 and 100 AU. The larger disks do not reach steady state in the timescale considered here, while the smaller disks have reached their collisional equilibrium by $\sim 1$ Gyr. 

A fit to the model population gives the relation $f \propto t^{-0.47}$ while a fit to the observed population yields $f \propto t^{-0.18 \pm 0.10}$, just within 3 $\sigma$ of the best-fit model. Fitting of the observed disks is strongly affected by the low number of disks in our data set, especially by the relatively few young disks ($<1$ Gyr) in the sample, but one can conclude from \Fig{fig:f_age} that the model population is a reasonable fit for the observed population. The validity of these fits will benefit from additional observations, which will allow more significant conclusions to be drawn with larger samples. Interesting to notice is the absence of a peak in excess flux analogous to the one seen in the observed A star samples around 10-15 Myr \citep{currie08}.

\subsection{Radius vs. age}

\begin{figure}
  \centering
  \includegraphics[width=8cm, angle=0]{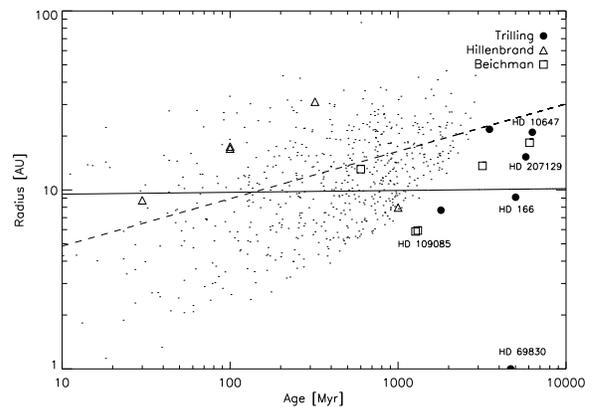}
  \caption{Disk radius plotted as a function of age. The model population is shown with small dots, and data symbols are the same as for \Fig{fig:F24}.  Only the populations which could be detected at both 24 and 70 $\micron$ are plotted. Also shown are power-law fits to the model (dashed line) and observed (solid line) populations. These are not in agreement, and possible reasons for this are discussed in the text. \label{fig:rad_age}}
\end{figure}

\Fig{fig:rad_age} is a plot of disk radius as a function of age, for those disks that could be detected at $24$ and $70\micron$. 

A lower limit for detectable radii is visible in the model population, with this lower limit increasing with age beyond $\sim$100 Myr. Older disks must therefore have larger radii to be detected at both $24$ and $70\micron$. This comes from the fact that disks with larger radii are slower in processing their mass and can therefore remain above detection thresholds for a longer time than disks with small radii. This radius increase with age is therefore only due to detection sensitivity considerations. In fact, there is as yet no evidence for a correlation between radius and age \citep{najitawilliams05}. 

A fit to the model population confirms the apparent increase of disk radius with age, with the relation $r \propto t^{0.25}$, while a fit to the observed population yields $r \propto t^{0.00 \pm 0.04}$. The fit to the data is strongly influenced by the disks from the \cite{hillenbrand08} sample, which consists of younger, and mostly larger disks. We also see a lack of observed small and young disks, which are predicted by the model. Although this could be evidence for rapid inner clearing, the small number of disks of the observed sample makes it difficult to draw robust conclusions as to the cause of this. Thus we note that it is worth scrutinising $<$300 Myr stars to see if there really is an absence of $<$10 AU disks that would be the precursors of older systems such as HD 40136 and HD 109085 ($\eta$ Corvi), and whether the radii derived from 24-70 $\micron$ colour temperatures appear smaller than reality for a significant number older disks, perhaps due to the presence of both hot and cold dust (as is known to be the case for $\eta$ Corvi; see \citealt{wyattgreaves05}), or due to the action of P-R drag (discussed later). In comparison, for sources detected at $70\micron$, \cite{carpenter09} fitted FEPS IRS spectra to derive probability distributions for inner disk radii peaking at a few tens of AU. 

The observed disks HD 69830 and HD 166 \citep{trilling08} appear on the plot where the model predicts that no disks should be detected (bottom right corner of \Fig{fig:rad_age}. For HD 69830, we attribute this to an unusually high disk luminosity; indeed this disk is found to have over 800 times the maximum theoretical luminosity for its radius, age and the spectral type of their host, as was already mentioned in \Sec{sec:f_rad}. HD 166 lies just outside the model population. Its disk is small for its age and its luminosity is just over its maximum theoretical luminosity; the unusual brightness for a disk of this age and colour temperature may mean that the disk around HD 166 has unusually strong or massive planetesimals compared to its peers. Other disks are also, although to a lesser extent, anomalously bright for their age: HD 109085, HD 10647 and HD 207129. Unusual disk properties might explain this: HD 109085 has two disk components, as discussed above, while recent imaging of the disk around HD 207129 \citep{krist10} and HD 10647 (Stapelfeldt et al., in preparation) with the ACS coronagraph on the \textit{Hubble Space Telescope} shows that these disks have radii larger than the 24-70$\micron$ radii used here by an order of magnitude. We discussed in \Sec{sec:effective} why we might expect blackbody radii to be smaller than the true radii, and these discrepancies also indicate that the 24-70$\micron$ radius could be a worse estimator for the true radius for FGK stars, compared to A stars. This could be because some of the disks have two components, as discussed in \Sec{sec:timeevolution}, or because there is a large spread in the $X_{r, F}$ factors between the disks. Larger disk samples and detailed analysis should help improve our understanding of this effect.

Dust might also find itself closer to the star than the planetesimal belt under the action of Poynting-Robertson drag. This becomes important when P-R drag and collisional timescales are comparable, or in terms of luminosity, when $f < \fpr$, with $\fpr$ being given by \Eq{eq:fpr}.

This limit is shown on \Fig{fig:f_rad}. From this plot, it emerges that disks with small radii could be affected by PR drag, given the proximity of the limit for PR drag to the limit of detection at $24\micron$ for these disks. In our sample, no system has $f/\fpr < 1$, meaning that PR drag is not significant for the observed disks we are considering ; the most likely system to be affected by this effect is HD 40136.  We also note that if the disk radius has been underestimated by a factor $X_r$, then the limit at which P-R drag becomes important given in \Eq{eq:fpr} should be lower by a factor $X_r^{-1/2}$. Therefore we do not expect P-R drag to have significantly influenced the blackbody radius, derived from dust temperature, which is thus a good proxy for the location of the planetesimal belt (with the caveats mentioned in \Sec{sec:effective}).

\subsection{Histogram of $f/\fmax$}

\begin{figure}
  \centering
  \includegraphics[width=8cm, angle=0]{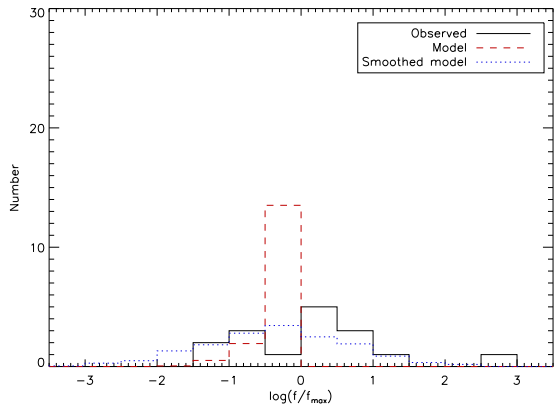}
  \caption{Histogram of dust fractional luminosity for the observed population with an excess at both 24 and 70 $\micron$ (solid line), the model population (dashed, red) and the model population with a Gaussian smoothing of 1 dex (dotted, blue). Luminostiies are expressed as a fraction of the maximum theoretical luminosity $\fmax$, given by \Eq{eq:fmaxfull}, and only the model population which could be detected at both 24 and 70 $\micron$ is plotted. \label{fig:hist_fmax}}
\end{figure}

\Fig{fig:hist_fmax} shows a histogram of the quantity $f/\fmax$, plotted for our data sets and for the model. We also plot the model with an arbitrary Gaussian smoothing of 1 dex to account for variations of disk properties between individual systems, which we have not considered in our model. The range of fluxes is broader in the data, which, as stated by \cite{wyatt07Astars} can be explained by the fact that we have assumed in our model that all the planetesimal belts have the same properties, where in reality the values of parameters such as $D_c$, $Q^*_D$ and $e$ are expected to vary from one belt to another. The parameters we used would then be correct as an average over a whole population of stars, but not necessarily for individual debris disks. This is particularly relevant when the sample of observed disks is as small as it is here. Although our data set is small, main features are reproduced by the smoothed model.

\section{Conclusions}\label{sec:conclusion}

In this paper, we have showed that the main features seen in observations of debris disks around FGK stars can be attributed to collisional grinding of planetesimals. We modelled data collected by teams with the Spitzer Space Telescope, and used 24 and 70$\micron$ colours to derive simple properties of the disks for each of these systems. Our results are consistent with models of planetesimal strengths and their size distributions, although we find parameters corresponding to these properties to be an order of magnitude different from those found in the study of A stars that was done using the same model \citep{wyatt07Astars}. 

We discussed whether these values might be effective values due to differences in stellar environment, and found that properties of disks around later-type stars may be somewhat different from those around earlier-type stars, because they evolve on longer timescales and therefore have had more time to form larger planetesimals (e.g. \citealt{kennedykenyon09}), which are then also stronger if the strength is mainly due to gravitational pull. We also find that disks around FGK stars are less massive than those around A stars by a small factor, in agreement with previous findings (e.g. \citealt{natta04}). Future observations to increase the sample of debris disks detected at multiple wavelengths will allow us to test our assumption that the blackbody disk radii underestimate the true radii of disks around FGK stars by a larger factor than for those around A stars. Realistic grain modelling analogous to the work done for A stars by \cite{bonsorwyatt10} could also help constrain this factor. 

We proposed two explanations for the discrepancy between the slower time evolution we derive in this analysis compared to that found by \cite{carpenter09} and \cite{siegler07}. We first suggested that this could be due to disks having two components. The samples of disks observed in these surveys are made up in large part of young ($<1$ Gyr) stars, around which the hot component would still be bright enough to produce detectable amounts of $24\micron$ emission. However, this excess emission would evolve rapidly due to the disk component's small distance from its host star. In contrast, the data we use in this analysis includes many older stars, for which the $24\micron$ emission would come from the cold component of the disk, which would evolve on longer timescales. However, an alternative explanation for the apparent slow evolution might also be that it comes from cold dust, with the excess emission past a few Myr then caused by transient events. 

Small data samples are strong caveats on these results. In most cases, the flux statistics we derive from our analysis are based on only a few disks. This shows the importance of collecting more observations of disks around late-type stars with large-scale surveys. More observations would help determine why the $24-70\micron$ radii might be worse radius estimators around stars of different types, and whether the slow evolution of $24\micron$ flux statistics we derived in this paper is statistically significant.

\section*{Acknowledgments}

NK acknowledges STFC studentship PPA/S/S/2006/04497 and an ESO Post-doctoral Fellowship, and thanks Grant Kennedy for useful discussions as well as the anonymous referee for helpful comments and suggestions.

\bibliographystyle{mn2e}
\bibliography{../thesisbib}
\bsp

\label{lastpage}

\end{document}